\documentclass{jetpl}
\twocolumn

\lat

\title{Study of 2$\beta$-decay of $^{100}$Mo and $^{82}$Se using the NEMO3 detector}

\rtitle{Study of 2$\beta$-decay of $^{100}$Mo and $^{82}$Se using the NEMO3 detector
\ldots}

\sodtitle{Study of 2$\beta$-decay of $^{100}$Mo and $^{82}$Se using the NEMO3 detector}

\author{R.~Arnold$^j$, C.~Augier$^h$, J.~Baker$^e$,
 A.~Barabash$^g~$\thanks{Corresponding author,
Institute of Theoretical and Experimental Physics, B.~Cheremushkinskaya 25,
117218 Moscow, Russia,  e-mail: Alexander.Barabash@itep.ru,
tel.: 007 (095) 129-94-68, fax: 007 (095) 883-96-01},
V.~Brudanin$^c$, A.J.~Caffrey$^e$, V.~Egorov$^c$,
J.L.~Guyonnet$^j$, F.~Hubert$^a$, Ph.~Hubert$^a$,
L.~Jenner$^n$, C.~Jollet$^j$, S.~Jullian$^h$, A.~Klimenko$^c$,
O.~Kochetov$^c$, S.~Konovalov$^g$, V.~Kovalenko$^c$,
D.~Lalanne$^h$, F.~Leccia$^a$, I.~Linck$^j$, C.~Longuemare$^b$, G.~Lutter$^a$,
Ch.~Marquet$^a$, F.~Mauger$^b$, H.W.~Nicholson$^i$,
H.~Ohsumi$^m$, F.~Piquemal$^a$, J-L.~Reyss$^d$,
R.~Saakyan$^n$, X.~Sarazin$^h$, F.~Scheibling$^j$, Yu.~Shitov$^c$,
L.~Simard$^h$, A.~Smolnikov$^c$, I.~\v{S}tekl$^k$,
J.~Suhonen$^f$, C.S.~Sutton$^i$, G.~Szklarz$^h$,
V.~Timkin$^c$, J.~Thomas$^n$, V.~Tretyak$^c$,
V.~Umatov$^g$, L.~V\'{a}la$^k$, I.~Vanyushin$^g$,
S.~Vasiliev$^c$, V.~Vasilyev$^g$,V.~Vorobel$^l$,
Ts.~Vylov$^c$}

\rauthor{R.\,Arnold et al.}

\sodauthor{R. Arnold et al.}

\address{$^a$CENBG, IN2P3-CNRS et Universit\'e de Bordeaux,
               33170 Gradignan, France}
\address{$^b$LPC, IN2P3-CNRS et Universit\'e de Caen, 14032 Caen, France}
\address{$^c$JINR, 141980 Dubna, Russia }
\address{$^d$CFR, CNRS, 91190 Gif sur Yvette, France}
\address{$^e$INEEL, Idaho Falls, ID 83415, U.S.A.}
\address{$^f$JYV\"ASKYL\"A University, 40351 Jyv\"askyl\"a, Finland}
\address{$^g$ITEP, 117218  Moscow, Russia}
\address{$^h$LAL, IN2P3-CNRS et Universit\'e Paris-Sud, 91405 Orsay, France}
\address{$^i$MHC, South Hadley, Massachusetts 01075, U.S.A.}
\address{$^j$IReS, IN2P3-CNRS et Universit\'e
              Louis Pasteur, 67037 Strasbourg, France.}
\address{$^k$IEAP CTU, Prague,  Czech Republic.}
\address{$^l$Charles University, Prague, Czech Republic.}
\address{$^m$SAGA University, Saga, Saga 840-8502, Japan}
\address{$^n$University College London, London, UK }

\dates{30 July 2004}{30 July 2004}

\abstract{After analysis of 5797 h of data from the detector NEMO3,  
new limits on neutrinoless  double beta decay of $^{100}$Mo 
($T_{1/2} > 3.1\cdot 10^{23}$ y , 90\% CL)  and $^{82}$Se ($T_{1/2} > 1.4\cdot 
10^{23}$ y , 90\% CL) have been obtained. The corresponding limits  
on the effective majorana neutrino mass are: $<m_{\nu}>$ $< (0.8-1.2)$ eV and 
$<m_{\nu}>$ $< (1.5-3.1)$ eV, respectively. Also the limits on double-beta decay with Majoron 
emission are: $T_{1/2} > 1.4\cdot 10^{22}$ y (90\%CL) for $^{100}$Mo 
and $T_{1/2}> 1.2\cdot 10^{22}$ y (90\%CL) for $^{82}$Se. Corresponding bounds on the 
Majoron-neutrino coupling constant are $<g_{ee}>$ $< (0.5-0.9) \cdot 10^{-4}$ 
and $< (0.7-1.6) \cdot 10^{-4} $. Two-neutrino $2\beta$-decay half-lives have 
been measured with a high accuracy, $T_{1/2}^{^{100}Mo} = [7.68 \pm 0.02(stat) \pm 
0.54(syst) ] \cdot 10^{18}$ y  and  $T_{1/2}^{^{82}Se} = [10.3 \pm 0.3(stat) \pm 
0.7(syst) ]\cdot 10^{19}$ y.}


\begin{document}

\maketitle

Interest in neutrinoless double beta decay ($2\beta0\nu$) has seen a significant rebirth  
in recent years after evidence for neutrino oscillations was obtained 
from the results of atmospheric \cite{link1}  and solar 
\cite{link2,link3,link4,link5,link6} neutrino experiments (see, for example, the 
discussions in \cite{link7,link8,link9}).

  This observation of oscillations was recently confirmed by the KamLAND experiment with reactor 
antineutrinos \cite{link10} and by the new SNO result \cite{link11}. These
results are an impressive proof  
that neutrinos have a non-zero mass. However, the experiments 
studying neutrino oscillations are not sensitive to the nature of the neutrino mass 
(Dirac or Majorana?) and provide no information on the absolute scale of the neutrino 
masses, since such experiments are sensitive only to the difference, $\Delta m^{2}$. The 
detection 
and study of $2\beta0\nu$ decay may clarify the following problems of neutrino physics 
(see discussions in \cite{link12,link13,link14}): (i) neutrino nature; is the neutrino a 
Dirac or a Majorana particle?, (ii) absolute neutrino mass scale (a measurement or a limit 
on m$_1$), (iii) the type of neutrino mass hierarchy (normal, inverted, or  quasidegenerate),
 (iv) CP violation in the lepton sector (measurement of the Majorana CP-violating phases).

     The main goal of the NEMO3 experiment is to study neutrinoless double beta decay of 
different isotopes ($^{100}$Mo, $^{82}$Se etc.) with a sensitivity of up to $\sim 10^{25}$ y,
 which corresponds to a sensitivity for the effective Majorana neutrino mass at the level 
of $\sim(0.1-0.3)$ eV \cite{link15}. The planned sensitivity for double beta decay with 
Majoron emission is $\sim 10^{23}$ y (the sensitivity for the coupling constant of Majoron to 
neutrino $<g_{ee}>$  is at the order of $\sim 10^{-5}$). In addition, one of the goals is a precise study of 
$2\beta2\nu$ decay for a number of nuclei ($^{100}$Mo, $^{82}$Se, $^{116}$Cd, $^{150}$Nd, 
$^{130}$Te, $^{96}$Zr and $^{48}$Ca) with high statistics and to study all major characteristics of 
the decay.

     NEMO3 is a tracking detector, which in contrast to $^{76}$Ge experiments 
\cite{link16,link17}, detects not only the total energy deposition but also other important 
parameters of the process. These include the energy of individual electrons, their angular 
distribution, the event vertex coordinates in the source plane, etc. This  provides 
a unique opportunity to monitor and reject backgrounds. Since June of 2002 NEMO3 has 
been running in the Frejus Underground Laboratory (France) located at a depth of 4800 m.w.e. 

     The detector has a cylindrical shape and consists of 20 identical sectors 
(see Fig~\ref{fig1}). A thin ($\sim$30-60 mg/cm$^2$) source is placed in the center 
of the detector which contains 2$\beta$-decaying nuclei and has a total area of 20 m$^2$ and 
a mass of about 10 kg. In particular, it includes 7.1 kg of enriched Mo (average 
enrichment is 98\%, the total mass of $^{100}$Mo is 6.914 kg) and 0.962 kg of Se (enrichment is 
97\%, the total mass of $^{82}$Se is 0.932 kg). To investigate the external background 
a part of the source is made of very pure natural materials (TeO$_2$ - 0.767 kg and Cu - 0.621 kg). 
The level of contamination of the sources with radioactive impurities was obtained from  
measurements using low-background HPGe-detectors.

      The basic detection principles of NEMO3 are the following. The energy of the electrons is 
measured with plastic scintillators coupled to PMTs (1940 individual counters), while 
the tracks are reconstructed from the information obtained with drift Geiger cells (6180 cells). 
The tracking volume of the detector is filled with a mixture of $\sim$95\% He, 4\% 
alcohol, 1\% Ar and 0.15\% water at 20 mbar above atmospheric pressure. In addition, a magnetic field of 
25 Gauss parallel to the detector's axis is created by a solenoid surrounding the detector. 
The magnetic field is used to identify electron-positron pairs to suppress the 
background associated with these events.

     The main characteristics of the detector's performance are the following. The energy 
resolution of the scintillation counters lies in the interval of 14-17\% (FWHM for 1 MeV 
electrons). The time resolution is 250 ps for an electron energy of 1 MeV. The reconstruction 
accuracy of a two electron (2e)  vertex is around 1 cm. The characteristics of the detector are 
studied in special calibration runs with radioactive sources. The energy 
calibration is carried out using $^{207}$Bi with conversion electrons at energies 
0.482 and 0.976 MeV, and $^{90}$Sr with the end-point of the $\beta$ spectrum at 2.283 MeV. 
The vertex reconstruction accuracy for 2e$^-$ events was determined in measurements with 
$^{207}$Bi. The time-of-flight properties were determined via measurements with $^{60}$Co where 
two $\gamma$-quanta are emitted simultaneously, $^{207}$Bi for which two electrons are 
emitted simultaneously and neutron sources which provide high energy electrons crossing the detector.

     The detector is surrounded by a passive shield made of 20 cm of steel, 30 cm of 
water contained in tanks covering the vertical exterior of the detector and wood and paraffin on the top and bottom. The level of radioactive impurities in the construction materials of the 
detector and the passive shield was tested with low-background HPGe detectors.

     From June to December of 2002 a number of calibration and test measurements were carried 
out, as was the first data taken for double beta decay studies. Since 14 
February 2003 the detector has been routinely taking double beta decay data . The calibrations 
with the radioactive sources are carried out every 1.5 months. Calorimeter 
stability is checked daily using a laser system between calibrations.
     A detailed description of the detector and its characteristics is presented 
in \cite{link18}.

     In this paper we present the results for $^{100}$Mo and $^{82}$Se. Only part 
(5797 h of measurement) of the available data has been analyzed.

\underline{\textbf{ Double beta decay of $^{100}$Mo.}}  	
2e-events with a common vertex at the source have been selected. For 2e-events, 
an electron  was defined as a track between the source foil and a fired scintillation counter with 
the energy deposited being greater than 200 keV. The track curvature must be consistent 
with a negatively charged particle. The time-of-flight measurement should be consistent 
with the hypothesis of two electrons leaving the source from a common vertex 
simultaneously. In order to suppress the $^{214}$Bi background, which is followed by a 
$^{214}$Po $\alpha$-decay, it is required that there are no delayed Geiger cell hits 
close to the event vertex or the electron track (within a delay of up to 700 $\mu$s). A typical 
2e-event is shown in Fig~\ref{fig2}.

	Fig~\ref{fig3} shows the $2\beta2\nu$ energy spectrum for $^{100}$Mo. The total 
number of useful events (after background subtraction) is $\sim$141000. The signal-to-background ratio is 40/1, while it is 100/1 for energies above 1 MeV. This means that 
the background is negligible. The detection efficiencies which included the selection 
cuts were estimated by Monte Carlo (MC) simulations.  This was done for two models of the 
$^{100}$Mo decay. The first one is the higher state dominance (HSD) mechanism and 
the second is the single state dominance (SSD) mechanism (see ref. \cite{link19}). 
For the SSD mechanism it is assumed that the decay goes via the lowest 1$^+$ state of 
$^{100}$Tc \cite{link20}. The detection efficiencies calculated by MC simulations were 5.68\% for the
SSD mechanism and 6.33\% for the HSD mechanism. Correspondingly, the following results were 
obtained for the $^{100}$Mo half-life:

$$
  T_{1/2} = [7.68 \pm 0.02(stat) \pm 0.54(syst) ]\cdot 10^{18}\; y\;  \textrm{(SSD\;)}       
$$

$$
  T_{1/2} = [8.57 \pm 0.02(stat) \pm 0.6(syst) ]\cdot 10^{18}\; y\;    
\textrm{(HSD\;)}       
$$

	A preliminary analysis of the experimental single electron energy spectrum favors 
the SSD hypothesis. However, at the moment this question remains open and requires further 
studies. Both values are in agreement 
with the world mean value from other experiments: $(8 \pm 0.7)\cdot 10^{18}$ y 
\cite{link21,link22}. 

\underline{$2\beta0\nu$-decay.}
	The energy range 2.65-3.2 MeV has been investigated. The information about the electron's energies and their angular distribution in 
2e-events can be used to improve the selection of candidate events and thus,  improve 
the NEMO3 sensitivity \cite{link24}. Introducing a higher threshold of 900 keV for each 
electron, one gets 13 candidate events in this energy window. This is in agreement with our 
expectations from the background (18.8 events) coming mainly from $^{222}$Rn (13 events) plus the $2\beta2\nu$ 
contribution (5.8 events). The calculated 
$2\beta0\nu$ efficiency is 8.2\%.   As a result a  
limit has been set at $T_{1/2} > 3.1 \cdot 10^{23}$  y (90\% CL), which is better than the best 
previous measurement ($T_{1/2} > 5.5 \cdot 10^{22}$  y \cite{link23}).
Using the nuclear 
matrix elements (NME) from \cite{link25,link26,civ03}, the following range of limits on the neutrino mass 
has been derived:   $<m_{\nu}>$ $< (0.8-1.2)$ eV. This range can be compared with that of the 
$^{76}$Ge experiment \cite{link16} (using the same nuclear matrix elements 
\cite{link25,link26,civ03}), $<m_{\nu}>$ $< (0.33-0.84)$ eV. 

     For neutrinoless double beta decay caused by the right current admixture ( $\lambda$ 
term) in weak interactions the 2.8-3.2 MeV energy window was used. An 
additional cut on the electron energies, $|E_1-E_2|>800$ keV \cite{link24}, allows an extra 
background suppression. In this case there are only 3 candidate events in the 2.8-3.2 MeV 
window with 2.8 (2.0+0.8) expected. With an efficiency of 4.2\% this gives $T_{1/2} > 1.8 \cdot 
10^{23}$  y (90\% CL) and $<\lambda>$ $< (1.5-2.0) \cdot 10^{-6}$, using the NME from \cite{Suh94}.

\underline{$2\beta0\nu\chi$ decay.}
	The energy interval 2.65-3.2 MeV has been studied. Introducing  
a threshold of 750 keV for each electron one gets only 18 events (13.7+8 are expected with an 
efficiency of 0.49\%) and the half-life limit yields $T_{1/2} > 1.4 \cdot 10^{22}$  y (90\% CL). 
This value is better than the previous one ($T_{1/2} > 5.8 \cdot 10^{21}$  y \cite{link27}) 
and the corresponding limit on the Majoron-neutrino coupling constant (NME from 
\cite{link25,link26,civ03}) is  $<g_{ee}>$  $< (0.5-0.9)\cdot 10^{-4}$ . This is one of the best 
limits for the majoron coupling constant.

\underline{\textbf{Double beta decay of $^{82}$Se.}}
 The energy spectrum of  $2\beta2\nu$-events for $^{82}$Se is shown in Fig~\ref{fig4}. A higher threshold of 300 keV was used in order to improve the signal-to-background ratio.
The total number of useful events after the background subtraction is $\sim1800$. The 
signal-to-background ratio is approximately 4 to 1. The detection efficiency has been calculated by 
MC calculations to be 6.02\%. The $^{82}$Se half-life value obtained is,
$$
     T_{1/2} = [10.3 \pm 0.3(stat) \pm 0.7(syst) ] \cdot 10^{19}  y.		               
$$

     This value is in agreement with our previous measurement with NEMO2 \cite{link28} and 
with the world average value of $(9 \pm 1) \cdot 10^{19}$  y [22]. 

\underline{$2\beta0\nu$-decay.}  
 Three events have been detected in the interval 2.65-3.2 MeV. The expected background in this energy windows is 5.3 events. Again this background is  mainly due to $^{222}$Rn.  
Efficiency of 15.8\% gives a limit for the $^{82}$Se 
0$\nu$-decay of $T_{1/2} > 1.4\cdot 10^{23}$  y (90\% CL), which is better than the 
previous value 
by one order of magnitude ($T_{1/2} > 1.4\cdot 10^{22}$  y at 90\% CL \cite{link29}). 
The corresponding limit on the effective neutrino mass, using NME from
\cite{link25,link26,civ03} is m$_{\nu}$ $< (1.5-3.1)$ eV.

     For neutrinoless double beta decay caused by the right current admixture ( $\lambda$ 
term) in weak interactions, the 2.65-3.2 MeV energy window was used and a cut on the
electron energy, $|E_1-E_2|>800$ keV \cite{link24} was applied. The number of events is zero and the
efficiency is 8.5\%. This gives $T_{1/2} > 1.1 \cdot 10^{23}$  y (90\% CL) and 
$<\lambda>$ $< (3.2-3.8) \cdot 10^{-6}$, using the NME from \cite{Tom91,suh91}.

\underline{$2\beta0\nu\chi$-decay.} 
	The energy interval 2.3-3.2 MeV was studied. The number of selected 
events is 39, while 40 is the estimated background. Using the efficiency estimated by 
MC calculations (4.2\%) the following limit was obtained, $T_{1/2} > 1.2 \cdot 
10^{22}$  y. (90\% CL). This is better than the previous best limit ($T_{1/2} > 2.4 
\cdot 10^{21}$  y \cite{link28}). Using this result  and the NME from \cite{link25,link26,civ03} 
one gets the corresponding limit for the Majoron-neutrino coupling constant $<g_{ee}>$ $ < (0.7-1.6) \cdot 10^{-4}$.

Presently, the NEMO3 tracking detector is continuing to collect data. At the 
same time, work is in progress to improve the detector's performance, background conditions 
and data analysis techniques. In particular a radon barrier tent has been installed and a 
radon trapping factory will be in operation in autumn of 2004, which will help reduce the 
$^{222}$Rn background by a factor of 10-100.  It is believed that the sensitivity of the experiment 
can be increased significantly in the near future. 

     The authors would like to thank the Modane Underground Laboratory staff for their 
technical assistance in running the experiment. Portions of this work were supported 
by a grant from INTAS N 03051-3431 and a grant NATO PST.CLG.980022.

\begin{figure*}
\begin{center}
\caption{Fig.1. The NEMO3 detector. 
1 - source foil; 2 - plastic
scintillator; 3 - low radioactive PMT; 4 - tracking chamber (6,180 octagonal
Geiger cells).
}
\label{fig1}
\end{center}
\end{figure*}

\begin{figure*}
\begin{center}
\caption{Fig.2. A view of a reconstructed 2e-event in NEMO3. Sum energy of the electrons is 2024 keV, energy of electrons in the pair is 961 keV and 1063 keV.
}
\label{fig2}
\end{center}
\end{figure*}

\begin{figure*}
\begin{center}
\caption{Fig.3. $2\beta2\nu$ spectrum of $^{100}$Mo, background subtracted.
}
\label{fig3}
\end{center}
\end{figure*}

\begin{figure*}
\begin{center}
\caption{Fig.4. $2\beta2\nu$ spectrum of $^{82}$Se, background subtracted.
}
\label{fig4}
\end{center}
\end{figure*}


\begin{thebibliography}{99}

\bibitem{link1}
 H. Sobel et al., Nucl. Phys. B (Proc. Suppl.) {\bf 91}, 127 (2001).

\bibitem{link2}
 B.T. Cleveland et al., Astrophys. J. {\bf 496}, 505 (1998).

\bibitem{link3}
 S. Fukuda et al., Phys. Rev. Lett. {\bf 86}, 5655 (2001); {\bf 86}, 5656 (2001).

\bibitem{link4}
 V.N. Gavrin et al., Nucl. Phys. B (Proc. Suppl.) {\bf 91}, 36 (2001).

\bibitem{link5}
 M. Altman et al., Phys. Lett. B {\bf 490}, 16 (2000).

\bibitem{link6}
 Q.R. Ahmad et all., Phys. Rev. Lett. {\bf 89}, 011301 (2002); {\bf 89},  011302 (2002). 

\bibitem{link7}
 P.C. de Holanda, A.Yu. Smirnov, Phys. Rev. D {\bf 66}, 113005 (2002). 

\bibitem{link8}
 J.N. Bahcall et al., JHEP {\bf 0207}, 054 (2002); hep-ph/0204314.

\bibitem{link9}
 M. Maltoni et al., Phys. Rev. D {\bf 67}, 013011 (2003); hep-ph/0207227.

\bibitem{link10}
 K. Eguchi et al., Phys. Rev. Lett. {\bf 90}, 021802 (2003).

\bibitem{link11}
 S.N. Ahmed et al., Phys. Rev. Lett. {\bf 92}, 181301 (2004).

\bibitem{link12}
S.M. Bilenky, hep-ph/0403245.


\bibitem{link13}
 S. Pascoli, S.T. Petcov, W. Rodejohann, Phys. Lett., B {\bf 558}, 141 (2003). 

\bibitem{link14}
 S. Pascoli, S.T. Petcov, Phys. Lett., B {\bf 580}, 280 (2004).

\bibitem{link15}
 NEMO3 Proposal, preprint 94-29, LAL Orsay, (1994).

\bibitem{link16}
 H.V. Klapdor-Kleingrothaus et al., Eur. Phys. J. A {\bf 12}, 147 (2001).

\bibitem{link17}
 C.E. Aalseth et al., Phys. Rev. D {\bf 65}, 092007 (2002).

\bibitem{link18}
 R. Arnold et al., preprint LAL 04-05, February 2004; hep-ph/0402115.

\bibitem{link19}
 F. Simkovic, P. Domin, S. Semenov, J. Phys. G {\bf 27}, 2233 (2001).

\bibitem{link20}
 J. Abad et al., Journal of Physique, {\bf 45}, 147 (1984); Ann. Fis. A {\bf 80}, 9 (1984). 

\bibitem{link21}
V.D. Ashitkov et al., JETP Lett., {\bf 76}, 529 (2001).

\bibitem{link22}
 A.S. Barabash, Czech. J. Phys. {\bf 52}, 567 (2002).


\bibitem{link23}
 H. Ejiri et al., Phys. Rev. C {\bf 63}, 065501 (2001).

\bibitem{link24}
 R. Arnold et al., Nucl. Instr. Meth. A {\bf 503}, 649 (2003).

\bibitem{link25}
 F. Simkovic et al., Phys. Rev. C {\bf 60}, 055502 (1999).

\bibitem{link26}
 S. Stoica, H.V. Klapdor-Kleingrothaus, Nucl. Phys. A {\bf 694}, 269 (2001).

\bibitem{civ03}
 O. Civatarese, J. Suhonen, Nucl. Phys. A {\bf 729}, 867 (2003).
\bibitem{Suh94}
 J. Suhonen, O. Civitarese, Phys. Rev. C {\bf 49}, 3055 (1994).

\bibitem{link27}
 H. Ejiri et al., Phys. Lett. B {\bf 531}, 190 (2002).


\bibitem{link28}
 R. Arnold et al., Nucl. Phys. A {\bf 636}, 209 (1998). 

\bibitem{link29}
 S.R. Elliott et al., Phys. Rev. C {\bf 46}, 1535 (1992).

\bibitem{Tom91}
 T. Tomoda, Rep. Prog. Phys. {\bf 54}, 53 (1991).

\bibitem{suh91}
 J. Suhonen, S.B. Khadkikar, A. Faessler, Nucl. Phys. A {\bf 535}, 509 (1991).
 


\end{thebibliography}
\end{document}